\documentclass[10pt]{article}
\usepackage{hyperref,amssymb,tikz-cd,float,graphicx}
\usepackage[utf8]{inputenc}
\usepackage[fleqn]{amsmath}
\usepackage{geometry}
\newcommand{\bs}[1]{\boldsymbol{#1}}
\newcommand{\qt}[1]{\overline{#1}}
\newcommand{\pq}[1]{\hat{#1}}
\newcommand{\dq}[1]{\tilde{#1}}
\newcommand{\pd}[1]{\check{#1}}

\begin{document}
\numberwithin{equation}{section}

\title{Dual Quaternion Variational Integrator for Rigid Body Dynamic Simulation}
\author{Xu, Jiafeng\\NTNU in Aalesund, Norway\\jiafeng.xu@ntnu.no \and Halse, Karl Henning\\NTNU in Aalesund, Norway\\karl.h.halse@ntnu.no}
\date{\vspace{-5ex}}
\maketitle
\begin{abstract}
We introduce a symplectic dual quaternion variational integrator(DQVI) for simulating single rigid body motion in all six degrees of freedom. Dual quaternion is used to represent rigid body kinematics and one-step Lie group variational integrator is used to conserve the geometric structure, energy and momentum of the system during the simulation. The combination of these two becomes the first Lie group variational integrator for rigid body simulation without decoupling translations and rotations. Newton-Raphson method is used to solve the recursive dynamic equation. This method is suitable for real-time rigid body simulations with high precision under large time step. DQVI respects the symplectic structure of the system with excellent long-term conservation of geometry structure, momentum and energy. It also allows the reference point and 6-by-6 inertia matrix to be arbitrarily defined, which is very convenient for a variety of engineering problems.
\end{abstract}
\section{Introduction}
In computer simulations, quaternions are commonly used for representing rotations in \textit{Special Orthogonal Group} \(SO(3)\) due to its clarity and compactness. Dual quaternions are an extension of the quaternion concept that also includes translations in the formulation. In recent years, the dual quaternion representation of rigid body kinematics has gained popularity in many fields, e.g., biomechanics\cite{dual_bio_pennestri}, cybernetics\cite{dual_descent_unsik}, robotics\cite{dq_robotics_alba,dual_dynamics_dooley} and computer graphics\cite{dual_blending_kavan}. The advantages dual quaternions have over many other formulations can be summarized as 1) Singularity-free; 2) Un-ambiguous; 3) Shortest path interpolation; 4) Most efficient and compact form (8-by-1 vector); 5) Unified representation of translation and rotation in a single invariant coordinate frame\cite{beginner_dual_quaternion}; 6) Intuitive connection of its exponential map to the screw motion.

On the other hand, geometric mechanics is a branch of mathematics that in principle applies geometric methods for systems whose configuration space is a Lie group\cite{geo_mecha_holm}. Variational integrators are powerful tools for geometric mechanics that aim at discretizing the formulation of a continuous system, instead of using differential equations. They are symplectic methods that preserve the geometric structure of the system and exactly conserve energy and momentum during system evolution, which is especially important for conservative systems. Matthew West discussed the algebraic property and engineering application of variational integrators in his in-depth Ph.D thesis\cite{variational_integrators_west}. Lie group variational integrator is an exceptionally efficient method that makes the system evolve automatically on Lie groups without the use of re-projection, constraints and local coordinates\cite{lie_integrator_overview_leok}. Many mathematical papers have introduced Euler-Poincaré and Lie-Poisson equations for a generic yet abstract variational approach for geometric mechanics\cite{euler_lie_mar,var_moscow_hernan,bracket_bloch}. However, in the subject of using Lie Group variational integrator for single rigid body dynamics, 3-by-3 rotation matrix is still widely used, also the translation and rotation of the rigid body are decoupled that only \(SO(3)\) and its Lie algebra \(\mathfrak{so}(3)\) are effectively used\cite{lie_group_matrix_underwater,lie_3d_pendulum_taeyoung,lie_integrator_full_lee}. 

This paper is inspired by Manchester and Peck's work\cite{quaternion_variational_zachary}, who for the first time used quaternions to represent rotation in Lie group variational integrator. Having noticed that the \textit{Special Euclidean Group} \(SE(3)\) for spatial displacement of rigid bodies(referred as \textit{poses}\cite{SE3_tutorial} in this paper) is a Lie Group by itself, we introduce the Dual Quaternion Variational Integrator(DQVI), a combination of Lie group variational integrator and the unit dual quaternion group, which is a 2-to-1 homomorphism to \(SE(3)\). To the best of our knowledge, this is the first simulation practice of Lie group variational integrator for rigid body dynamics without decoupling the translation and rotation. DQVI allows both the reference point and the full 6-by-6 inertia matrix to be arbitrarily defined, which is very convenient for a variety of engineering problems such as the implementation of added mass and damping coefficients matrices in the simulation of water-borne vehicles.

Section 2 reviews the background of the algebraic and kinematic properties of quaternions and dual quaternions. Section 3 explains the relationship of the exponential maps of unit quaternions and unit dual quaternions with the screw motion in the context of Lie group theory. Section 4 formulates the discretized dynamic equations for single rigid body, which will be used in the variational integrator. Section 5 focuses on solving the dynamic equations using Newton-Raphson method. Section 6 is the conclusion.
\section{Background}
\subsection{Quaternions}
\textit{Quaternions} is a four-dimensional vector space over the real numbers with one real element and three imaginary elements. It was first described by Hamilton\cite{letter_quaternion_hamilton} and denoted as \(\mathbb{H}\) for his honor\footnote[1]{In general, the notation system for quaternions and dual quaternions has not been standardized yet, readers may find discrepancies in this paper from other works}. In this paper all quaternions are marked by an overline \(\overline{\phantom{q}}\)
\[
    \qt{\bs{q}} \in \mathbb{H} \simeq \mathbb{R}^4;\quad \qt{\bs{q}} =  q_w+q_x\bs{i}+q_y\bs{j}+q_z\bs{k}; \quad \bs{i}^2=\bs{j}^2=\bs{k}^2=\bs{ijk}=-1
\]
The quaternions can be written in vector form as
\[
    \qt{\bs{q}} = \begin{bmatrix}q_w\\\bs{q}_{xyz}\end{bmatrix}; \quad \bs{q}_{xyz} = \begin{bmatrix}q_x\\q_y\\q_z\end{bmatrix}
\]
Based on its definition, the quaternions in their vector form has the following unique operators
\begin{itemize}
    \item Multiplication
    \[\qt{\bs{q}}_1 \circ \qt{\bs{q}}_2 = 
    \begin{bmatrix}q_{1w}q_{2w}-\bs{q}_{1n}\cdot \bs{q}_{2xyz}\\
    q_{1w}\bs{q}_{2xyz} + q_{2w}\bs{q}_{1xyz}+ \bs{q}_{1xyz}\times\bs{q}_{2xyz}\end{bmatrix}\]
    \item Conjugate
    \[\overline{\bs{q}}^\dagger=\begin{bmatrix}q_w\\ -\bs{q}_{xyz}\end{bmatrix}\]
\end{itemize}
The multiplication of quaternions is not commutative. Also there is a useful algebraic property about the quaternion multiplication and the vector dot product
\begin{equation}\label{eq:quaternion_dot_product_rule}
    (\qt{\bs{q}}_1\circ \qt{\bs{q}}_2)\cdot \qt{\bs{q}}_3 = (\qt{\bs{q}}_1^\dagger\circ \qt{\bs{q}}_3)\cdot \qt{\bs{q}}_2 = (\qt{\bs{q}}_3\circ \qt{\bs{q}}_2^\dagger)\cdot \qt{\bs{q}}_1
\end{equation}
\subsection{Unit Quaternion for SO(3)}
The \textit{Unit Quaternion}, denoted as \(\mathbb{H}_u\), is a subset of quaternions with unit length, which topologically forms a \textit{3-sphere} \(\mathbb{S}^3\) in \(\mathbb{R}^4\).
\[\mathbb{H}_u = \{\qt{\bs{q}}\in \mathbb{R}^4||\qt{\bs{q}}|=1\}\]
\(\mathbb{H}_u\) is an non-abelian Lie group under the quaternion multiplication \(\circ :\mathbb{H}_u \times\mathbb{H}_u \rightarrow \mathbb{H}_u\). The inverse is its conjugate \(\overline{\bs{q}}^\dagger\) and the identity is
\[\qt{\bs{I}} =\qt{\bs{q}}\circ \qt{\bs{q}}^\dagger = \qt{\bs{q}}^\dagger \circ \qt{\bs{q}} = \begin{bmatrix}1\\0\\0\\0\end{bmatrix}\]

The rotation group \(SO(3)\) is equivalent to the \textit{real projective space} \(\mathbb{RP}^3\) of the antipodal point pairs on \(\mathbb{S}^3\). Hence the map \(\varphi :\mathbb{H}_u \rightarrow SO(3)\) defined by \(\varphi(\qt{\bs{q}}) = \{\pm \qt{\bs{q}}\}\) is a 2-to-1 homomorphism\cite{naive_lie_theory}.

A vector \(\bs{r}\in \mathbb{R}^3\) is represented as a \textit{Pure Quaternion} whose ``real part" equals zero. In this paper all pure quaternions are denoted with a hat \(\pq{\bs{\phantom{q}}}\)
\[\pq{\bs{r}} = \begin{bmatrix}0\\\bs{r}\end{bmatrix}\]
the rotation of \(\bs{r}\) can be achieved by a combination of two quaternion rotations in \(\mathbb{S}^3\).
\begin{equation}\label{eq:unit_quaternion_rotation}
    \qt{\bs{q}} = \begin{bmatrix}\cos{\frac{\theta}{2}}\\ \bs{n}\sin{\frac{\theta}{2}}\end{bmatrix}; \quad \pq{\bs{r}}' = \qt{\bs{q}}\circ \pq{\bs{r}}\circ \qt{\bs{q}}^\dagger
\end{equation}
\(\bs{n}\) is the rotating axis, which is the eigenvector of 3-by-3 rotation matrix. It has identical coordinates in two coordinate systems defined by the rotation. In rigid dynamics they are usually referred as \textit{world-fixed frame} \(\bs{e}_W\) and \textit{body-fixed frame} \(\bs{e}_B\). \(\theta\) is the angle of rotation. The rotation around \(\bs{n}\) with angle \(\theta\) is equivalent to a rotation around \(-\bs{n}\) with angle \(-\theta\).
\subsection{Dual Quaternion}
The \textit{Dual Quaternion}, denoted as \(\mathbb{DH}\), is a Clifford algebra comprised of two quaternions denoted as \(\qt{\bs{p}}_a\) and \(\qt{\bs{p}}_b\) for its real and dual part respectively.  In this paper all dual quaternions are marked by a tilde \(\dq{\phantom{q}}\)
\[
    \dq{\bs{p}}\in \mathbb{DH}\simeq \mathbb{R}^8; \quad \dq{\bs{p}} = \qt{\bs{p}}_a + \epsilon \qt{\bs{p}}_b
\]
\(\epsilon\) is the \textit{dual unit} that resembles the imaginary unit in complex number but with its square as zero
\[
    \epsilon^2 = 0; \quad\epsilon \neq 0
\]
The concept was firstly proposed by Clifford\cite{biquaternions_clifford} for representing vectors with not only magnitude and direction, but also position, which is naturally associated with rigid body poses. Based on its definition, the dual quaternions in their vector form also have the following unique operators 
\begin{itemize}
    \item Multiplication
    \[\dq{\bs{p}}_1\otimes\dq{\bs{p}}_2 = \begin{bmatrix}\qt{\bs{p}}_{1a} \circ \qt{\bs{p}}_{2a}\\\qt{\bs{p}}_{1a}\circ \qt{\bs{p}}_{2b}+\qt{\bs{p}}_{1b}\circ \qt{\bs{p}}_{2a}\end{bmatrix}\]
    \item Quaternion Conjugate
    \[\dq{\bs{p}}^\dagger = \begin{bmatrix}\qt{\bs{p}}_a^\dagger\\ \qt{\bs{p}}_b^\dagger\end{bmatrix}\]
    \item Dual Transpose
    \[\dq{\bs{p}}^*=\begin{bmatrix}\qt{\bs{p}}_b\\ \qt{\bs{p}}_a\end{bmatrix}\]
\end{itemize}
The multiplication of dual quaternions is also not commutative. Based on Equation \ref{eq:quaternion_dot_product_rule}, there is also an algebraic property of the dual quaternion multiplication and the vector dot product
\begin{equation}\label{eq:dual_quaternion_dot_product_rule}
    (\dq{\bs{p}}_1\otimes \dq{\bs{p}}_2)\cdot\dq{\bs{p}}_3 = (\dq{\bs{p}}_1^\dagger\otimes \dq{\bs{p}}_3^*)\cdot \dq{\bs{p}}_2^* = (\dq{\bs{p}}_3^*\otimes \dq{\bs{p}}_2^\dagger)\cdot \dq{\bs{p}}_1^*
\end{equation}

The \textit{Pure Dual Quaternion} is the dual quaternion composed by two pure quaternions, the representation of a vector \([\bs{r}_a,\bs{r}_b]^T\in \mathbb{R}^6\) in pure dual quaternion form is denoted with a check \(\pd{\bs{\phantom{p}}}\)
\[
    \pd{\bs{r}} = \begin{bmatrix}\pq{\bs{r}}_a \\ 
    \pq{\bs{r}}_b\end{bmatrix}
\]
\subsection{Unit Dual Quaternion for SE(3)}
The \textit{Unit Dual Quaternion}, denoted as \(\mathbb{DH}_u\), is a subset of dual quaternion that forms a Lie group\cite{dh_bedia} with identity
\[
    \dq{\bs{I}} = \dq{\bs{p}}\otimes\dq{\bs{p}}^\dagger = \dq{\bs{p}}^\dagger\otimes\dq{\bs{p}} = \qt{\bs{I}} +\epsilon\pq{\bs{0}}
\]
It can be used to represent poses in \(SE(3)\), which is the semidirect product of the translation group and the rotation group \(SE(3) = \mathbb{R}^3 \rtimes SO(3)\). If the translation displacement is denoted as \(\bs{l}\in \mathbb{R}^3\) and the rotation is represented by unit quaternion \(\varphi (\qt{\bs{q}}) = SO(3)\), the unit dual quaternion representation of a pose \(\psi(\dq{\bs{p}}) = SE(3)\) can be formed as
\begin{equation}\label{eq:unit_dual_quaternion}
    \dq{\bs{p}} = \qt{\bs{q}} + \epsilon\frac{1}{2}\pq{\bs{l}}\circ \qt{\bs{q}}
\end{equation}
The map \(\psi : \mathbb{DH}_u \rightarrow SE(3)\) is also a 2-to-1 homomorphism \(\psi (\dq{\bs{p}}) = \{\pm \dq{\bs{p}}\}\)
\[
    \psi (\dq{\bs{p}}_1)\times \psi (\dq{\bs{p}}_2) = \psi (\dq{\bs{p}}_1\otimes\dq{\bs{p}}_2)
\]

A pose \(\dq{\bs{p}}\) can also be used to transform a point vector \(\bs{r}\in \mathbb{R}^3\) in \(SE(3)\). Construct a pose with identity \(\mathbb{H}_u\)
\[
    \dq{\bs{r}} = \qt{\bs{I}} +\epsilon\frac{\pq{\bs{r}}}{2}
\]
then the new point vector \(\bs{r}'\) after transformation is
\begin{equation}
    \dq{\bs{r}}' = \dq{\bs{p}}\otimes\dq{\bs{r}}; \quad \pq{\bs{r}}' = 2\qt{\bs{r}}'_b\circ {\qt{\bs{r}}'_a}^\dagger
\end{equation}
The transformation also applies to nested poses that are commonly used in multibody kinematics.
\begin{equation}
    \dq{\bs{r}}' = \dq{\bs{p}}_1\otimes\dq{\bs{p}}_2\otimes \dq{\bs{r}}
\end{equation}
\subsection{Translational and Angular Velocity}
The translational and angular velocities are essential to the rigid body simulation. They are denoted as \(\bs{v}_B\) and \(\bs{\omega}_B\) in \(\bs{e}_B\), \(\bs{v}_W\) and \(\bs{\omega}_W\) in \(\bs{e}_W\). Use the notations in Equation \ref{eq:unit_dual_quaternion}, \(\bs{v}_B\) has the following relation with the pose
\begin{equation}\label{eq:v_B_quaternion}
    \bs{v}_W = \dot{\bs{l}}; \quad \pq{\bs{v}}_B = \qt{\bs{q}}^\dagger\circ \pq{\bs{v}}_W \circ \qt{\bs{q}}
\end{equation}
Also for \(\bs{\omega}_B\), if a vector \(\bs{r}_B\) is constant in \(\bs{e}_B\), its form \(\bs{r}_W\) in \(\bs{e}_W\) is
\begin{equation}
    \pq{\bs{r}}_W= \qt{\bs{q}}\circ \pq{\bs{r}}_B\circ \qt{\bs{q}}^\dagger
\end{equation}
The time derivative of \(\bs{r}_W\) has the relation
\[
    \dot{\pq{\bs{r}}}_W = \dot{\qt{\bs{q}}}\circ \pq{\bs{r}}_B\circ \qt{\bs{q}}^\dagger + \qt{\bs{q}}\circ \pq{\bs{r}}_B\circ \dot{\qt{\bs{q}}}^\dagger; \quad
    \dot{\bs{r}}_W = 2(\dot{\qt{\bs{q}}}\circ \qt{\bs{q}})_n\times \bs{r}_W = \bs{\omega}_W\times \bs{r}_W
\]
Hence
\begin{equation}
    \pq{\bs{\omega}}_W = 2\dot{\qt{\bs{q}}}\circ \qt{\bs{q}}^\dagger; \quad \pq{\bs{\omega}}_B = 2\qt{\bs{q}}^\dagger\circ \dot{\qt{\bs{q}}} 
\end{equation}
Construct a 6-by-1 vector \(\bs{\chi}\) with \(\bs{v}\) and \(\bs{\omega}\)
\begin{equation}\label{eq:dual_velocity}
    \bs{\chi}_W = \begin{bmatrix}\bs{\omega}_W\\\bs{v}_W\end{bmatrix}; \quad
    \bs{\chi}_B = \begin{bmatrix}\bs{\omega}_B\\\bs{v}_B\end{bmatrix}
\end{equation}
Its pure dual quaternion form has a key relation with the pose
\begin{equation}\label{eq:dual_velocity_quaternion}
    \pd{\bs{\chi}}_W = 2\dot{\dq{\bs{p}}}\otimes \dq{\bs{p}}^\dagger; \quad \pd{\bs{\chi}}_B = 2\dq{\bs{p}}^\dagger\otimes \dot{\dq{\bs{p}}}
\end{equation}
\section{Exponential Map and Screw Motion}
\subsection{Exponential Map on SO(3)}
The Lie algebra of \(SO(3)\) is denoted as \(\mathfrak{so}(3)\). They are connected by exponential map
\[
    exp: \mathfrak{so}(3)\rightarrow SO(3)
\]
The exponential of a quaternion by definition has the expression
\begin{equation}
    exp\left(\qt{\bs{\eta}}\right) = \sum_{k=1}^{\infty}\frac{\qt{\bs{\eta}}^k}{k!}=exp\left(\eta_w\right)\cdot
    \begin{bmatrix}
    \cos{|\bs{\eta}_{xyz}|}\\
    sin{|\bs{\eta}_{xyz}|}\cdot\bs{\eta}_{xyz}\left/|\bs{\eta}_{xyz}|\right.
    \end{bmatrix}
\end{equation}
To have a unit quaternion homomorphic to \(SO(3)\), \(\qt{\bs{\eta}}\) must be a pure quaternion, i.e. the exponential of a pure quaternion is a unit quaternion. So the exponential map on \(\mathbb{H}_u\) is denoted as
\begin{equation}
    exp : \mathfrak{H}_u\rightarrow \mathbb{H}_u; \quad
    \mathfrak{H}_u = \{\pq{\bs{\eta}}\in \mathbb{H}|\bs{\eta}\in \mathbb{R}^3\}    
\end{equation}
\(\mathfrak{H}_u\) is also the Lie algebra of \(\mathbb{H}_u\).
Equation \ref{eq:unit_quaternion_rotation} can now be written in a compact parametrized form
\begin{equation}
    \qt{\bs{q}} = exp\left(\pq{\bs{n}}\cdot \frac{\theta}{2}\right)
\end{equation}
\subsection{Exponential Map on SE(3) and Screw Motion}
\(SE(3)\) as a Lie group also has its Lie algebra \(\mathfrak{se}(3)\). Homomorphically \(\mathbb{DH}_u\)  also has its exponential map
\begin{equation}
    exp: \mathfrak{DH}_u\rightarrow \mathbb{DH}_u; \quad \mathfrak{DH}_u = \{\pd{\bs{\eta}}\in \mathbb{DH}_u|\bs{\eta}\in \mathbb{R}^6\}
\end{equation}
Similar to the exponential map of \(\mathbb{H}_u\), the exponential map of \(\mathbb{DH}_u\) employs the dual angle \(\Theta = \theta_a + \epsilon\theta_b\) and pure dual quaternions \(\pd{\bs{s}} = \pq{\bs{s}}_a+\epsilon\pq{\bs{s}}_b\). 
\begin{equation}
    \dq{\bs{p}} = exp\left(\pd{\bs{s}}\cdot \frac{\Theta}{2}\right)
\end{equation}The geometric interpretation of those four quantities is related to \textit{screw motion}, i.e. a rotation and a translation about the same axis. According to Chasle's theorm\cite{chasles_theorm_daniilidis}, any rigid transformation can be described by a screw motion. \(\theta_a\) is the angle of rotation, unit vector \(\bs{s}_a\) is the direction of the axis of rotation, \(\theta_b\) is the magnitude of translation along the axis and \(\bs{s}_b\) is the \textit{moment of the axis}, which is given by equation \(\bs{s}_b = \bs{k}\times \bs{s}_a\). \(\bs{k}\) is the position vector of a point that the axis passes, the exact position of the point is irrelevant since \(\bs{s}_b\) will remain constant. 

The resemblance in the exponential maps of quaternions and dual quaternions reveals that quaternion is only a special case of dual quaternion where the rotation axes pass through the origin, while dual quaternion can represent rotations with arbitrary axes.

Finally, the relationship of all groups can be represented as follows
\begin{equation}
\begin{tikzcd}[column sep=6pc]
\mathfrak{H}_u \arrow{r}{d\varphi} \arrow{d}{exp} & 
  \mathfrak{so}(3) \arrow{d}{exp} & \mathfrak{se}(3)\arrow{d}{exp}&\mathfrak{DH}_u\arrow{d}{exp}\arrow[swap]{l}{d\psi}\\
  \mathbb{H}_u \arrow{r}{\varphi} & SO(3) \arrow{r}{\mathbb{R}^3\rtimes}&SE(3)&\mathbb{DH}_u \arrow[swap]{l}{\psi}
\end{tikzcd}
\end{equation}
\section{Dynamic Equations for Single Rigid Body}
\subsection{Continuous Formulation}
The equations for a single rigid body motion satisfies the integral Lagrange d'Alembert's principle
\begin{equation}\label{eq:d'alembert_principle}
    \delta \int_{t_0}^{t_1}L\left(\dq{\bs{p}},\dot{\dq{\bs{p}}},t\right)dt+\int_{t_0}^{t_1}\dq{\bs{F}}\left(\dq{\bs{p}},\dot{\dq{\bs{p}}},t\right)\cdot\delta \dq{\bs{p}}dt=0
\end{equation}
where \(L\) is the Lagrangian of the system described by a pose and its derivative. \(\dq{\bs{F}}\in \mathbb{R}^8\) is the generalized non-conservative force also expressed as dual quaternion. \(\delta \dq{\bs{p}}\) is the infinitesimal variation of pose that vanishes at the end point but otherwise arbitrary. This of course leads to the Euler-Lagrange equations
\begin{equation}\label{eq:euler_lagrange}
    \frac{\partial L}{\partial \dq{\bs{p}}} - \frac{d}{dt}\frac{\partial L}{\partial \dot{\dq{\bs{p}}}}+\dq{\bs{F}}=0 
\end{equation}
The kinetic energy \(T\) of a single rigid body is
\begin{equation}\label{eq:kinetic_energy}
    T = \frac{1}{2}\int_m\left(\bs{v}_B+\bs{\omega}_B\times \bs{\rho}\right)^2dm=\frac{1}{2}\bs{\chi}_B\cdot\bs{M}_{6\times 6}\cdot \bs{\chi}_B
\end{equation}
where \(\bs{\rho}\) is the position vector of a mass point to the reference point in \(\bs{e}_B\) and 
\begin{equation}\label{eq:6x6_M}
    \bs{M}_{6\times 6} =
    \begin{bmatrix}
    \bs{J}_{3\times 3}& m\cdot S(\bs{r}_g) \\
    -m\cdot S(\bs{r}_g)&\bs{m}_{3\times 3} 
    \end{bmatrix}
\end{equation}
\(\bs{J}_{3\times 3}\) is the moment of inertia tensor with respect to the reference point. \(m\) is the mass of the rigid body and \(\bs{m}_{3\times 3}\) is the diagonal mass matrix. Usually the mass on three directions are identical, but for hydrodynamic related problems, those three elements may be different due to the added mass. 
\(\bs{r}_g\) is the position vector of center of mass of the rigid body in body-fixed frame and \(S(\cdot)\) is the skew-symmetric matrix of cross product operator
\[
    S(\bs{r}) =
    \begin{bmatrix}
    0&-r_z&r_y\\
    r_z&0&-r_x\\
    -r_y&r_x&0
    \end{bmatrix}; \quad
    S(\bs{r}_1)\cdot \bs{r}_2 = \bs{r}_1\times \bs{r}_2
\]
In the context of unit dual quaternion pose, Equation \ref{eq:6x6_M} can be extended into an \(8\times 8\) matrix with extra rows and columns filled with zeros.
\begin{equation}\label{eq:8x8_M}
    \bs{M}_{8\times 8} =
    \begin{bmatrix}
    0&\cdots&0&\cdots\\
    \vdots&\bs{J}_{3\times 3}&\vdots& m\cdot S(\bs{r}_g) \\
    0&\cdots&0&\cdots\\
    \vdots&-m\cdot S(\bs{r}_g)&\vdots& \bs{m}_{3\times 3}
    \end{bmatrix}
\end{equation}
Use Equation \ref{eq:dual_velocity_quaternion}, Equation \ref{eq:6x6_M} can be written with unit dual quaternions
\begin{equation}\label{eq:continuous_kinetic}
    T = \frac{1}{2}\dq{\bs{\chi}}_B^T\cdot \bs{M}_{8\times 8}\cdot \dq{\bs{\chi}}_B = 2\left(\dq{\bs{p}}\otimes \dot{\dq{\bs{p}}}\right)^T\cdot \bs{M}_{8\times 8}\cdot \left(\dq{\bs{p}}\otimes \dot{\dq{\bs{p}}}\right)
\end{equation}
The major improvement here for kinetic energy formulation, compared with previous works\cite{lie_group_matrix_underwater,lie_3d_pendulum_taeyoung,lie_integrator_full_lee}, is that the reference point does not need to be situated at the center of mass and can now be arbitrarily defined. \(\bs{M}_{6\times 6}\) can now also be arbitrarily defined as long as the determinant is not zero. As a result, the potential energy of the single rigid body is a function of both translation and rotation of the rigid body \(U = U(\dq{\bs{p}})\).

The Lagrangian of the system is
\begin{equation}\label{eq:pose_L}
    L = T-U = 2\left(\dq{\bs{p}}^\dagger\otimes \dot{\dq{\bs{p}}}\right)^T\cdot \bs{M}_{8\times 8}\cdot \left(\dq{\bs{p}}^\dagger\otimes \dot{\dq{\bs{p}}}\right)-U\left(\dq{\bs{p}}\right)
\end{equation}
\subsection{Discrete Formulation}
Equation \ref{eq:d'alembert_principle} can be discretized into
\begin{equation}\label{eq:discrete_d'alembert_principle}
    \delta \sum_{k=0}^{N-1}L_{dk} + \sum_{k=0}^{N-1}W_{dk} = 0
\end{equation}
where \(L_{dk}\) and \(W_{dk}\) represent the action integral and virtual work integral respectively.
\subsubsection{Kinetic energy integral variation}
Equation \ref{eq:pose_L} shows that the single rigid body system is time invariant. \(\dot{\dq{\bs{p}}}_k\) can be approximated by trapezoidal rule with a fixed time step \(h = t_{k+1}-t_k\).
\begin{equation}\label{eq:discrete_assumption}
    \dot{\dq{\bs{p}}}_k = \left(\dq{\bs{p}}_{k+1}-\dq{\bs{p}}_k\right)\left/h\right.
\end{equation}
Higher order quadrature rules can also be applied, but in real-time simulations, usually there is an ``Update" function that overwrites data from last time frame. One-step quadrature rules work better for its simplicity and low cost. The integral of the kinetic energy can be represented as
\begin{equation}\label{eq:kinetic_energy_integral}
    T_{dk}\left(\dq{\bs{p}}_k,\dot{\dq{\bs{p}}}_k\right) \approx \frac{2}{h}\left(\dq{\bs{p}}_k^\dagger\otimes \left(\dq{\bs{p}}_{k+1}-\dq{\bs{p}}_k\right)\right)^T\cdot \bs{M}_{8\times 8}\cdot \left(\dq{\bs{p}}_k^\dagger\otimes \left(\dq{\bs{p}}_{k+1}-\dq{\bs{p}}_k\right)\right)
\end{equation}
Simplify the expression by using the fact that \(\dq{\bs{p}}_k^\dagger\otimes\dq{\bs{p}}_k = \dq{\bs{I}}\) and \(\bs{M}_{8\times 8}\cdot \dq{\bs{I}} = (\dq{\bs{I}}\cdot\bs{M}_{8\times 8})^T = \dq{\bs{0}}\)
\begin{equation}\label{eq:kinetic_energy_integral_simplified}
    T_{dk} = \frac{2}{h}\left(\dq{\bs{f}}_k^T\cdot \bs{M}_{8\times 8}\cdot \dq{\bs{f}}_k\right)
\end{equation}
where \(\dq{\bs{f}}_k = \dq{\bs{p}}_k^\dagger\otimes \dq{\bs{p}}_{k+1}\), so that \(\dq{\bs{p}}_{k+1} = \dq{\bs{p}}_k\otimes \dq{\bs{f}}_k\). By requiring that \(\dq{\bs{f}}_k \in \mathbb{DH}_u\), \(\dq{\bs{p}}_k\) is ensured to be evolving in \(\mathbb{DH}_u\) as well.
The variation of \(\dq{\bs{f}}_k\) is
\begin{equation}\label{eq:f_variation}
    \delta \dq{\bs{f}}_k = \delta \dq{\bs{p}}_k^\dagger\otimes \dq{\bs{p}}_{k+1} + \dq{\bs{p}}_k^\dagger\otimes \delta \dq{\bs{p}}_{k+1}=\dq{\bs{f}}_k\otimes \pd{\bs{\eta}}_{k+1}-\pd{\bs{\eta}}_k\otimes \dq{\bs{f}}_k
\end{equation}
where \(\pd{\bs{\eta}}_k \in \mathfrak{DH}_u\). 
So the variation of action integral can be written as
\begin{equation}\label{eq:kinetic_energy_integral_variation}
    \delta \sum_{k=0}^{N-1}T_{dk} = \sum_{k=0}^{N-1} \frac{4}{h}\dq{\bs{\zeta}}_k\cdot\left(\dq{\bs{f}}_k\otimes \pd{\bs{\eta}}_{k+1}-\pd{\bs{\eta}}_k\otimes \dq{\bs{f}}_k \right)
\end{equation}
where 
\[
    \dq{\bs{\zeta}}_k = \dq{\bs{f}}_k^T \cdot \bs{M}_{8\times 8}
\] 
Due to the structure of \(\bs{M}_{8\times 8}\), \(\dq{\bs{\zeta}}_k\) is guaranteed to be pure dual quaternion \(\pd{\bs{\zeta}}_k\). 
With some index manipulation
\begin{equation}\label{eq:kinetic_energy_integral_variation_indexed}
\begin{split}
    \delta \sum_{k=0}^{N-1}T_{dk} =& \frac{4}{h}\pd{\bs{\zeta}}_{N-1}\cdot\left(\dq{\bs{f}}_{N-1}\otimes\pd{\bs{\eta}}_N\right)-\frac{4}{h}\pd{\bs{\zeta}}_0\cdot\left(\pd{\bs{\eta}}_0\otimes\dq{\bs{f}}_0\right)\\
    &+\sum_{k=1}^{N-1}\frac{4}{h}\pd{\bs{\zeta}}_{k-1}\cdot\dq{\bs{f}}_{k-1}\otimes \pd{\bs{\eta}}_k-\pd{\bs{\zeta}}_k\cdot\pd{\bs{\eta}}_k\otimes \dq{\bs{f}}_k
\end{split}
\end{equation}
The variations are computed with the boundary points \(\dq{\bs{p}}_0\) and \(\dq{\bs{p}}_N\) held fixed, meaning \(\pd{\bs{\eta}}_k\) vanishes at \(k=0\) and \(k=N\)
\begin{equation}\label{eq:0N_eta_vanish}
    \pd{\bs{\eta}}_0 = \pd{\bs{\eta}}_N = \pd{\bs{0}}
\end{equation}
Also use Equation \ref{eq:dual_quaternion_dot_product_rule}, Equation \ref{eq:kinetic_energy_integral_variation_indexed} is then modified as
\begin{equation}\label{eq:kinetic_energy_integral_variation_final}
    \delta \sum_{k=0}^{N-1}T_{dk} = \sum_{k=1}^{N-1} \frac{4}{h}\left(\dq{\bs{f}}_{k-1}^\dagger\otimes\pd{\bs{\zeta}}_{k-1}^* -\pd{\bs{\zeta}}_k^*\otimes\dq{\bs{f}}_k^\dagger\right)\cdot \pd{\bs{\eta}}_k^*
\end{equation}
\subsubsection{Potential energy integral variation}
The potential energy is only the function of \(\dq{\bs{p}}\), so the potential energy integral variation can be represented as 
\begin{equation}\label{eq:potential_energy_integral_variation}
    \delta \sum_{k=0}^{N-1}U_{dk} \approx \sum_{k=0}^{N-1}\frac{h}{2}\frac{\partial U(\dq{\bs{p}}_k)}{\partial  \dq{\bs{p}}_k}\cdot \delta \dq{\bs{p}}_k+ \sum_{k=0}^{N-1}\frac{h}{2}\frac{\partial U(\dq{\bs{p}}_{k+1})}{\partial  \dq{\bs{p}}_{k+1}}\cdot \delta \dq{\bs{p}}_{k+1}
\end{equation}
After index manipulation and also using Equation \ref{eq:dual_quaternion_dot_product_rule} \& \ref{eq:0N_eta_vanish} , Equation \ref{eq:potential_energy_integral_variation} is simplified as 
\begin{equation}\label{eq:potential_energy_integral_variation_generalized}
    \delta \sum_{k=0}^{N-1}U_{dk} = \sum_{k=1}^{N-1}h\left(\dq{\bs{p}}_k^\dagger\otimes\left(\frac{\partial U(\dq{\bs{p}}_k)}{\partial  \dq{\bs{p}}_k}\right)^*\right)\cdot \pd{\bs{\eta}}_k^*
\end{equation}
\subsubsection{Virtual work integral variation}
The virtual work integral is approximated as
\begin{equation}\label{eq:virtual_work_integral_variation}
\begin{split}
    \sum_{k=0}^{N-1}W_d \approx&\sum_{k=0}^{N-1}\frac{h}{2}\left(\dq{\bs{F}}_k \delta \dq{\bs{p}}_k+ \dq{\bs{F}}_{k+1} \delta \dq{\bs{p}}_{k+1}\right)
\end{split}
\end{equation}
After index manipulation and also using Equation \ref{eq:dual_quaternion_dot_product_rule} \& \ref{eq:0N_eta_vanish}, Equation \ref{eq:virtual_work_integral_variation} is simplified as
\begin{equation}\label{eq:virtual_work_integral_variation_indexed}
    \sum_{k=0}^{N-1}W_{dk}=\sum_{k=1}^{N-1}h\left(\dq{\bs{F}}_k\cdot \left(\dq{\bs{p}}_k\otimes \pd{\bs{\eta}}_k\right)\right)= \sum_{k=1}^{N-1} h\left(\dq{\bs{p}}_k^\dagger\otimes \dq{\bs{F}}_k^*\right)\cdot \pd{\bs{\eta}}_k^*
\end{equation}
Alternatively, the generalized force in dual quaternion space can be transformed from classical torque and force in Cartesian frame by using the identity of work. Let \(\bs{\tau}_B\) and \(\bs{\tau}_W\) denote torque and force of all 6-DOF in \(\bs{e}_B\) and \(\bs{e}_W\) respectively.
\begin{equation}\label{eq:identity_work}
    W = \bs{\tau}_B\cdot \bs{\chi}_B = \bs{\tau}_W\cdot\bs{\chi}_W = \dq{\bs{F}}\cdot \dot{\dq{\bs{p}}}
\end{equation}
By using Equation \ref{eq:dual_quaternion_dot_product_rule} and \ref{eq:dual_velocity_quaternion} \begin{equation}\label{eq:dual_force_relation}
    \left(\dq{\bs{p}}^\dagger\otimes\dq{\bs{F}}^*\right)\cdot \pd{\bs{\chi}}_B^* = 2\pd{\bs{\tau}}_B^* \cdot \pd{\bs{\chi}}_B^*; \quad \left(\dq{\bs{F}}^*\otimes\dq{\bs{p}}\right)\cdot \pd{\bs{\chi}}_W^* = 2\pd{\bs{\tau}}_W^* \cdot \pd{\bs{\chi}}_W^*
\end{equation}
Since \(\pd{\bs{\chi}}_B\) and \(\pd{\bs{\chi}}_W\) are both arbitrary pure dual quaternions, the 2nd-4th and 6th-8th equations of
\begin{equation}
    \dq{\bs{F}}^*= 2\dq{\bs{p}}\otimes \pd{\bs{\tau}}_B^* = 2\pd{\bs{\tau}}_W^* \otimes \dq{\bs{p}}^\dagger
\end{equation}
are valid. Also, because \(\pd{\bs{\eta}}_k\) is a pure dual quaternion, Equation \ref{eq:virtual_work_integral_variation} can be simplified as 
\begin{equation}\label{eq:virtual_work_integral_variation_final}
    \sum_{k=0}^{N-1}W_{dk} = \sum_{k=1}^{N-1}2h\pd{\bs{\tau}}_{Bk\_external}^*\cdot \pd{\bs{\eta}}_k^*
\end{equation}
Notice that Equation \ref{eq:potential_energy_integral_variation_generalized} can also be simplified by using body-fixed conservative torque and force in Cartesian frame. The the 2nd-4th and 6th-8th equations of
\begin{equation}\label{eq:conventional_conservative_force}
    \pd{\bs{\tau}}_{Bk\_conservative}^* = -\frac{1}{2}\dq{\bs{p}}_k^\dagger\otimes\left(\frac{\partial U(\dq{\bs{p}}_k)}{\partial  \dq{\bs{p}}_k}\right)^*
\end{equation}
are valid. Hence
\begin{equation}\label{eq:potential_energy_integral_variation_final}
    \delta \sum_{k=0}^{N-1}U_{dk} = \sum_{k=1}^{N-1}2h\pd{\bs{\tau}}_{Bk\_conservative}^*\cdot \pd{\bs{\eta}}_k^*
\end{equation}
Return Equation \ref{eq:kinetic_energy_integral_variation_final}, \ref{eq:virtual_work_integral_variation_final} and \ref{eq:potential_energy_integral_variation_final} back to Equation \ref{eq:discrete_d'alembert_principle}
\begin{equation}\label{eq:discrete_eta_sum}
    \sum_{k=1}^{N-1}\left[\frac{4}{h}\left(\dq{\bs{f}}_{k-1}^\dagger\otimes\pd{\bs{\zeta}}_{k-1}^* -\pd{\bs{\zeta}}_k^*\otimes\dq{\bs{f}}_k^\dagger\right)+2h\left(\pd{\bs{\tau}}_{Bk\_conservative}^*+\pd{\bs{\tau}}_{Bk\_external}^*\right)\right]\cdot \pd{\bs{\eta}}_k^* = 0
\end{equation}
Since \(\pd{\bs{\eta}}_k\) are arbitrary perturbations in every time frame, Equation \ref{eq:discrete_eta_sum} must hold 
\begin{equation}\label{eq:discrete_dq}
    \dq{\bs{f}}_{k-1}^\dagger\otimes\pd{\bs{\zeta}}_{k-1}^* -\pd{\bs{\zeta}}_k^*\otimes\dq{\bs{f}}_k^\dagger+ \frac{h^2}{2}\left(\pd{\bs{\tau}}_{Bk\_conservative}^*+\pd{\bs{\tau}}_{Bk\_external}^*\right)=\begin{bmatrix}\lambda_1\\\bs{0}\\\lambda_2\\\bs{0}\end{bmatrix}
\end{equation}
\(\lambda_1\) and \(\lambda_2\) are unknown values, meaning Equation \ref{eq:discrete_dq} is only useful with its 2nd-4th and 6th-8th equations. Equation \ref{eq:discrete_dq} forms a recursive map for getting \(\dq{\bs{p}}_{k+1}\) from \(\dq{\bs{p}}_k\) and \(\dq{\bs{p}}_{k-1}\). The evolution of the system  is defined given the initial values of \(\dq{\bs{p}}_0\) and \(\dq{\bs{p}}_1\).
\subsection{Parametrization}
Since \(\dq{\bs{f}}_k \in \mathbb{DH}_u\) is homomorphic to \(SE(3)\), it can be parametrized by 6 independent variables. Based on the definition of \(\mathbb{DH}_u\) in Equation \ref{eq:unit_dual_quaternion} and an unconstrained vector representation of \(\mathbb{H}_u\) \cite{quaternion_variational_zachary}, \(\dq{\bs{f}}\) can be expressed as
\begin{equation}\label{eq:unit_dual_quaternion_parametrization}
    \qt{\bs{f}}_a = \begin{bmatrix}\sqrt{1-\bs{\Phi}^2}\\\bs{\Phi}\end{bmatrix}, \quad \qt{\bs{f}}_b =
    \begin{bmatrix}
        -\frac{\bs{\Psi}\cdot\bs{\Phi}}{\sqrt{1-\bs{\Phi}^2}}\\
        \bs{\Psi}
    \end{bmatrix}
\end{equation}
Other ways for parametrization also exists. Put Equation \ref{eq:unit_dual_quaternion_parametrization} back to Equation \ref{eq:discrete_dq} and only treat \(\bs{\Phi}_k\) and \(\bs{\Psi}_k\) as variables
\begin{equation}\label{eq:para_AB}
    \left\{\begin{matrix}
        \bs{\mathcal{A}}(\bs{\Phi}_k, \bs{\Psi}_k) - \bs{\alpha}_k = \bs{0}\\
        \bs{\mathcal{B}}(\bs{\Phi}_k, \bs{\Psi}_k) - \bs{\beta}_k = \bs{0}
    \end{matrix}\right.
\end{equation}
where
\begin{equation}
\begin{split}
    \bs{\mathcal{A}}(\bs{\Phi}_k, \bs{\Psi}_k) =&\left(-\frac{\bs{\Psi}_k\cdot\bs{\Phi}_k}{\sqrt{1-\bs{\Phi}_k^2}}\bs{I}+S(\bs{\Psi}_k)\right)\cdot\left(\bs{M}_{21}\bs{\Phi}_k+\bs{M}_{22}\bs{\Psi}_k\right)\\
    &+\left(\sqrt{1-\bs{\Phi}_k^2}\bs{I}+S(\bs{\Phi}_k)\right)\cdot\left(\bs{M}_{11}\bs{\Phi}_k+\bs{M}_{12}\bs{\Psi}_k\right)
\end{split}
\end{equation}
\begin{equation}
    \bs{\mathcal{B}}(\bs{\Phi}_k, \bs{\Psi}_k) =\left(\sqrt{1-\bs{\Phi}_k^2}\bs{I}+S(\bs{\Phi}_k)\right)\cdot\left(\bs{M}_{21}\bs{\Phi}_k+\bs{M}_{22}\bs{\Psi}_k\right)
\end{equation}
\begin{equation}
\begin{split}
    \bs{\alpha}_k =&\left(\sqrt{1-\bs{\Phi}_{k-1}^2}\bs{I}-S(\bs{\Phi}_{k-1})\right)\cdot \left(\bs{M}_{11}\bs{\Phi}_{k-1}+\bs{M}_{12}\bs{\Psi}_{k-1}\right)\\
    &+\left(-\frac{\bs{\Psi}_{k-1}\cdot\bs{\Phi}_{k-1}}{\sqrt{1-\bs{\Phi}_{k-1}^2}}\bs{I}-S(\bs{\Psi}_{k-1})\right)\cdot \left(\bs{M}_{21}\bs{\Phi}_{k-1}+\bs{M}_{22}\bs{\Psi}_{k-1}\right)\\
    &+\frac{h^2}{2}\left(\bs{\tau}_{ak\_conservative}+\bs{\tau}_{ak\_external}\right)
\end{split}
\end{equation}
\begin{equation}
\begin{split}
    \bs{\beta}_k = &\left(\sqrt{1-\bs{\Phi}_{k-1}^2}\bs{I}-S(\bs{\Phi}_{k-1})\right)\cdot\left(\bs{M}_{21}\bs{\Phi}_{k-1}+\bs{M}_{22}\bs{\Psi}_{k-1}\right)\\
    &+\frac{h^2}{2}\left(\bs{\tau}_{bk\_conservative}+\bs{\tau}_{bk\_external}\right)
\end{split}
\end{equation}
\subsection{Retrieving Velocity}
In practice, the translational and angular velocity in Cartesian frame are used more frequently than \(\bs{\Phi}_k, \bs{\Psi}_k\) in \(\dq{\bs{f}}_k\). 
The transformation between \(\bs{\Phi}_k, \bs{\Psi}_k\) and \(\bs{\chi}_{Bk}\) can be achieved from the identity of kinetic energy. 
\begin{equation}
    T_k=\frac{1}{2}\pd{\bs{\chi}}_{Bk}^T\cdot\bs{M}_{8\times 8}\cdot \pd{\bs{\chi}}_{Bk} = \frac{1}{2}{}^{D}\dq{\bs{P}}_k\cdot \dot{\dq{\bs{p}}}_k
\end{equation}
where \(\pd{\bs{P}}_{Dk}\) denote the momenta in dual quaternion frame.
Hence
\begin{equation}\label{eq:identity_kinetic}
    \left(\pd{\bs{\chi}}_{Bk}^T\cdot\bs{M}_{8\times 8}\right)\cdot \pd{\bs{\chi}}_{Bk} = \left(\dq{\bs{p}}_k^\dagger\otimes\pd{\bs{P}}_{Dk}^*\right)^*\cdot \pd{\bs{\chi}}_{Bk}
\end{equation}
Based on Equation \ref{eq:discrete_assumption}, the same trapezoidal approximation can be made for \(\pd{\bs{P}}_{Dk}\) in discrete form
\begin{equation}\label{eq:discrete_kinetic}
    \pd{\bs{P}}_{Dk} = \left(\frac{\partial T}{\partial \dot{\dq{\bs{p}}}}\right)_k\approx \frac{\partial T_{dk}}{\partial \dq{\bs{p}}_k}
\end{equation}
Equation \ref{eq:discrete_kinetic} shows that \(\pd{\bs{P}}_{Dk}\) is not a function of \(\pd{\bs{\chi}}_{Bk}\). Since \(\pd{\bs{\chi}}_{Bk}\) is a pure dual quaternion, after taking derivative of \(\pd{\bs{\chi}}_{Bk}\) on both side of Equation \ref{eq:identity_kinetic}, a simplification can be made that the 2nd-4th and 6th-8th equations of
\begin{equation}
    \bs{M}_{8\times 8}\cdot\pd{\bs{\chi}}_{Bk} =\frac{1}{2} \left(\dq{\bs{p}}^\dagger\otimes\pd{\bs{P}}_{Dk}^*\right)^* 
\end{equation}
are valid.
After tedious but straightforward simplification, a familiar formulation appears
\begin{equation}\label{eq:momentum_evolution_map}
    \bs{M}_{6\times 6}\cdot \bs{\chi}_{Bk} = \frac{2}{h}\cdot
    \begin{bmatrix}
    \bs{\mathcal{A}}(\bs{\Phi}_k, \bs{\Psi}_k) \\[1em]
    \bs{\mathcal{B}}(\bs{\Phi}_k, \bs{\Psi}_k)
    \end{bmatrix}
\end{equation}
Equation \ref{eq:momentum_evolution_map} reveals that Equation \ref{eq:para_AB} is actually the evolution map of momentum in \(\bs{e}_B\). And in each time frame with given \(\bs{\Phi}_k\) and \(\bs{\Psi}_k\), the translational and angular velocity in \(\bs{e}_B\) can be retrieved by
\begin{equation}\label{eq:retrieving_velocity}
    \bs{\chi}_k = \frac{2}{h}\cdot\bs{M}_{6\times 6}^{-1}\cdot \begin{bmatrix}
    \bs{\mathcal{A}}(\bs{\Phi}_k, \bs{\Psi}_k) \\[1em]
    \bs{\mathcal{B}}(\bs{\Phi}_k, \bs{\Psi}_k)
    \end{bmatrix}
\end{equation}
\section{Newton-Raphson Solver}
\subsection{Jacobian Matrix}
To solve Equation \ref{eq:para_AB}, Newton-Raphson method is used. The 6-by-6 Jacobian matrix is
\begin{equation}
    \bs{\mathcal{J}}(\bs{\Phi}_k, \bs{\Psi}_k)=
    \begin{bmatrix}
    \frac{\partial\bs{\mathcal{A}}_k}{\partial\bs{\Phi}_k}&\frac{\partial\bs{\mathcal{A}}_k}{\partial\bs{\Psi}_k}\\[1em]
    \frac{\partial\bs{\mathcal{B}}_k}{\partial\bs{\Phi}_k}&\frac{\partial\bs{\mathcal{B}}_k}{\partial\bs{\Psi}_k}
    \end{bmatrix}
\end{equation}
where
\begin{equation}
\begin{split}
    \frac{\partial\bs{\mathcal{A}}}{\partial\bs{\Phi}} =
    &(\bs{M}_{21}\bs{\Phi}+\bs{M}_{22}\bs{\Psi})\cdot\frac{(\bs{\Phi}\times (\bs{\Phi}\times\bs{\Psi})-\bs{\Psi})^T}{(1-\bs{\Phi}^2)^{\frac{3}{2}}}+\left(-\frac{\bs{\Psi}\cdot\bs{\Phi}}{\sqrt{1-\bs{\Phi}^2}}\bs{I}+S(\bs{\Psi})\right)\cdot \bs{M}_{21}\\
    &-\frac{\left(\bs{M}_{11}\bs{\Phi}+\bs{M}_{12}\bs{\Psi}\right)\cdot\bs{\Phi}^T}{\sqrt{1-\bs{\Phi}^2}}-S(\bs{M}_{11}\bs{\Phi}+\bs{M}_{12}\bs{\Psi})+\left(\sqrt{1-\bs{\Phi}^2}\bs{I}+S(\bs{\Phi})\right)\cdot \bs{M}_{11}
\end{split}
\end{equation}
\begin{equation}
\begin{split}
    \frac{\partial\bs{\mathcal{A}}}{\partial\bs{\Psi}} =&-\frac{(\bs{M}_{21}\bs{\Phi}+\bs{M}_{22}\bs{\Psi})\cdot\bs{\Phi}^T}{\sqrt{1-\bs{\Phi}^2}}-S(\bs{M}_{21}\bs{\Phi})\\
    &-\frac{\bs{\Psi}\cdot\bs{\Phi}}{\sqrt{1-\bs{\Phi}^2}}\bs{M}_{22}+\left(\sqrt{1-\bs{\Phi}^2}\bs{I}+S(\bs{\Phi})\right)\cdot \bs{M}_{12}
\end{split}
\end{equation}
\begin{equation}
    \frac{\partial\bs{\mathcal{B}}}{\partial\bs{\Phi}} =-\frac{(\bs{M}_{21}\bs{\Phi}+\bs{M}_{22}\bs{\Psi})\cdot\bs{\Phi}^T}{\sqrt{1-\bs{\Phi}^2}}-S(\bs{M}_{21}\bs{\Phi}+\bs{M}_{22}\bs{\Psi})+\left(\sqrt{1-\bs{\Phi}^2}\bs{I}+S(\bs{\Phi})\right)\cdot \bs{M}_{21}
\end{equation}
\begin{equation}
    \frac{\partial\bs{\mathcal{B}}}{\partial\bs{\Psi}} =\left(\sqrt{1-\bs{\Phi}^2}\bs{I}+S(\bs{\Phi})\right)\cdot \bs{M}_{22}
\end{equation}
If the reference point is chosen at the center of mass, \(\bs{\mathcal{J}}(\bs{\Phi}_k, \bs{\Psi}_k)\) can be simplified as
\begin{equation}
\begin{split}
    \frac{\partial\bs{\mathcal{A}}}{\partial\bs{\Phi}} =&(\bs{M}_{22}\bs{\Psi})\cdot\frac{(\bs{\Phi}\times (\bs{\Phi}\times\bs{\Psi})-\bs{\Psi})^T}{(1-\bs{\Phi}^2)^{\frac{3}{2}}}-\frac{\left(\bs{M}_{11}\bs{\Phi}\right)\cdot\bs{\Phi}^T}{\sqrt{1-\bs{\Phi}^2}}\\
    &-S(\bs{M}_{11}\bs{\Phi})+\left(\sqrt{1-\bs{\Phi}^2}\bs{I}+S(\bs{\Phi})\right)\cdot \bs{M}_{11}
\end{split}
\end{equation}
\begin{equation}
    \frac{\partial\bs{\mathcal{A}}}{\partial\bs{\Psi}} =-\frac{(\bs{M}_{22}\bs{\Psi})\cdot\bs{\Phi}^T}{\sqrt{1-\bs{\Phi}^2}}-\frac{\bs{\Psi}\cdot\bs{\Phi}}{\sqrt{1-\bs{\Phi}^2}}\bs{M}_{22}
\end{equation}
\begin{equation}
    \frac{\partial\bs{\mathcal{B}}}{\partial\bs{\Phi}} =-\frac{(\bs{M}_{22}\bs{\Psi})\cdot\bs{\Phi}^T}{\sqrt{1-\bs{\Phi}^2}}-S(\bs{M}_{22}\bs{\Psi})
\end{equation}
\begin{equation}
    \frac{\partial\bs{\mathcal{B}}}{\partial\bs{\Psi}}=\left(\sqrt{1-\bs{\Phi}^2}\bs{I}+S(\bs{\Phi})\right)\cdot \bs{M}_{22}
\end{equation}
\subsection{Initial Estimation}
In order to achieve faster convergence in Newton-Raphson solver, a proper starting point should be estimated in every time frame. The initial condition of the system is usually given as the pose-velocity or pose-momentum form. In either form, the initial pose and momentum are given.
The definition of \(\dq{\bs{f}}_k\) gives out that
\begin{equation}
    \dq{\bs{f}}_k = \dq{\bs{p}}_k^\dagger\otimes\dq{\bs{p}}_{k+1} \approx\dq{\bs{p}}_k^\dagger\otimes(\dq{\bs{p}}_k+h\dot{\dq{\bs{p}}}_k) = \dq{\bs{I}}+\frac{h}{2}\pd{\bs{\chi}}_{Bk}
\end{equation}
Hence
\begin{equation}\label{eq:estimation_f}
    \begin{bmatrix}
    \bs{\Phi}_k\\
    \bs{\Psi}_k
    \end{bmatrix}
    \approx\frac{h}{2}\bs{\chi}_{Bk}
\end{equation}
Equation \ref{eq:estimation_f} reflects a constraint in this parametrization that \(|\bs{\Phi}|< 1\), meaning the time step \(h\) must be chosen small enough to ensure that the incremental rotations between adjacent times steps are less than 180 degrees. Equation \ref{eq:estimation_f} also serves as the starting point for Newton-Raphson solver to get the initial value of \(\bs{\Phi}_0\) and \(\bs{\Psi}_0\). Then for every time frame, put \(\bs{\Phi}_{k-1}\) and \(\bs{\Psi}_{k-1}\) in Equation \ref{eq:para_AB} to find \(\bs{\mathcal{A}}_k\) \& \(\bs{\mathcal{B}}_k\). Finally, use Equation  \ref{eq:retrieving_velocity} \& \ref{eq:estimation_f} to get the 
starting point of solving Equation \ref{eq:para_AB}.
If the reference point is at the center of mass and there is no rotation, the first estimation of \(\bs{\Psi}_k\) will be the exact solution, meaning the Newton-Raphson solver for pure translations is equivalent to the one-step forward Euler method. An iteration time of 3 can already converge the solution to machine precision.
\section{Conclusion}
The Dual Quaternion Variational Integrator(DQVI) introduced in this paper treats the rigid body pose in \(SE(3)\) as an entire Lie group without decoupling translation and rotation. A parametrization with 6 independent variables is used to ensure the system evolves in \(SE(3)\). Compared with traditional 2nd-order Newton-Euler ODE, DQVI preserve the geometric structure, energy and momentum of the system.  DQVI also allows reference point and 6-by-6 inertia matrix to be arbitrarily defined, providing more genericity to possible applications. Multi-step integration method can be developed based on our work. More application examples with the addition of potential field and non-conservative force will be made in the future.

\bibliographystyle{plain}
\bibliography{short}
\end{document}